\documentclass{aastex}
\usepackage{spr-astr-addons}
\usepackage{url}
\usepackage{threeparttable}

\begin{document}

\title{A study of line widths and kinetic parameters of ions in the solar corona}


\author{G. Q. Zhao \altaffilmark{1}} \and \author{D. J. Wu}
\affil{Purple Mountain Observatory, CAS, Nanjing 210008, China}
\and
\author{C. B. Wang}
\affil{CAS Key Laboratory of Geospace Environment, School of Earth and Space Sciences, University of Science and Technology of China, Hefei 230026, China}

\altaffiltext{1}{University of Chinese Academy of Sciences, Beijing 100049, China}

\begin{abstract}
Solar extreme-ultraviolet (EUV) lines emitted by highly charged ions have been extensively studied to discuss the issue of coronal heating and solar wind acceleration. Based on observations of the polar corona by the SUMER/SOHO spectrometer, this paper investigates the relation between the line widths and kinetic parameters of ions. It is shown that there exists a strongly linear correlation between two variables $(\sigma/\lambda)^2$ and $M^{-1}$, where $\sigma$, $\lambda$ and $M$ are the half-width of the observed line profile at $1/\sqrt{e}$, the wavelength and the ion mass, respectively. The Pearson product-moment correlation coefficients exceed 0.9. This finding tends to suggest that the ions from a given height of polar corona have a common temperature and a common non-thermal velocity in terms of existing equation. The temperature and non-thermal velocity are obtained by linear least-square fit. The temperature is around $2.8$ MK at heights of 57$''$ and 102$''$. The non-thermal velocity is typical 21.6 km s$^{-1}$ at height of 57$''$ and 25.2 km s$^{-1}$ at height of 102$''$.
\end{abstract}


\section{Introduction} \label{s:intro}

The measurement of solar extreme-ultraviolet (EUV) lines is an important tool to study the outstanding problems of coronal heating and solar wind acceleration \citep[e.g.,][]{has90p77,erd07p26}. Previous EUV measurements showed that most solar EUV lines are broader than those expected from thermal broadening in terms of the Doppler effect, demonstrating the presence of unresolved mass motions of the plasma \citep[i.e. non-thermal motions in literatures;][]{bol73p61,bol75p97,dos76p17,mar92}. The non-thermal motions are usually regarded as a manifestation of some type of wave (or turbulence), such as magneto-acoustic wave or Alfv\'en wave \citep{bol73p61,has90p77,erd98p87,ban98p08}. These waves may carry necessary energy to larger heights and contribute to the coronal heating and solar wind acceleration \citep[see also,][]{ofm05p67,cra07p20,mci11p77,chm13p40}. The wave energy density, for a given plasma density, depends on the wave velocity amplitude represented by the non-thermal velocity. This non-thermal velocity, as well as the ion temperature, appears as a Doppler broadening of an emission line. Hence, one need to first determine the ion temperature and non-thermal velocity from the line width for a quantitative discussion of coronal heating and solar wind acceleration.

It is difficult to discriminate the ion temperature and non-thermal velocity. All difficulties arise since these two pieces of information merge in only one physically observable quantity of line width \citep{dol08p71}. The separation from line width into an ion temperature and a non-thermal velocity depends on certain assumption. One usually makes an assumption about one of the above quantities to obtain the other one, depending on the purpose of the author. For instance, to obtain the non-thermal velocity one may assume that the unknown ion temperature is equal to the formation temperature \citep{has90p77,ban09p15,liy09p29,coy11p15}. These assumptions may limit the accuracy of the determined results \citep{lan09p94,bem12p10}. \citet{see97p87} have questioned the validity of the frequently used assumption of ion temperature equal to a given formation temperature of the ion, generally taken as the temperature of maximum abundance of the ion emitting the line \citep{jor69p01,arn85p25,arn92p94}. The derived magnitude of non-thermal velocity is quite sensitive to the ion temperature value. Moreover, if the ion temperature were to increase with height, then the increase in temperature could lead to the line width broadening without necessarily invoking an increase in the non-thermal velocity \citep{sin06p75}.

It will become easy to determine the coronal ion temperature and non-thermal velocity for off-limb observations, if two preconditions are fulfilled. They are the ions within a volume element having a common non-thermal velocity and having a common temperature. The condition of a common non-thermal velocity makes sense and seems acceptable, since it is generally believed that this velocity comes from mass motions of the plasma due to wave or turbulence, i.e. regardless of the ion species. In contrast, one may question another condition of a common temperature, because the corona has a low plasma density and thus may become highly collisionless especially for the coronal hole plasma \citep{tuc98p75}. In addition, \citet{mor03p57} performed a test for a common ion temperature and non-thermal velocity using their spectral line observations. Their results showed that there is not a common non-thermal velocity if there is a common ion temperature, and there is not a common ion temperature if there is a common non-thermal velocity. It should be noted that the data used for the test in \citet{mor03p57} are summed within a range of coronal heights from 1.02 to 1.06 solar radius.

The present paper aims to report a finding of strong correlation between the line widths and ion masses for the line width data at two given coronal heights. The paper is organized as follows. We describe the data in Section 2 and show the strong correlation between the line widths and ion masses in Section 3. On the basis of this correlation, the coronal ion temperature and non-thermal velocity at the given coronal heights are deduced in Section 4. Finally, the conclusion and discussions are presented in Section 5.

\section{Observational data}

The SUMER instrument on board the $SOHO$ spacecraft provides authors with unprecedented opportunity for studying the solar atmosphere by analysing ion emission lines \citep{wil95p89,wil97p75,dom95p01,lem97p05}. \citet{dol08p71} have performed observations with a 1$''$ $\times$ 300$''$ slit, oriented in the north-south direction, using detector A. In particular, they obtained four data sets of line widths, and the main one, which is our interest, consists of much more ion species as shown later. The main data set is for a not well-developed polar coronal hole with possibility of a mixture of the quiet corona. For the main data set, the complete observation lasted for about 13 hours on 30 May 2002. The authors have used these data to discuss the issue of preferential ion heating.

\begin{table*}
\caption{Line width data at two positions above the solar limb \citep{dol08p71}.}
\centering
\begin{tabular}{l c l l}
\hline\hline
& & \multicolumn{2}{c}{Line width$^b$ (\AA)} \\ [-2.5ex]
Line$^a$ & atomic mass ($M_H$) \\ [-2.0ex]
& &  \multicolumn{1}{c}{57$''$} & \multicolumn{1}{c}{102$''$} \\ [0.0ex]
\hline
Fe \uppercase\expandafter{\romannumeral10} ~1028.02  & 55.85 & 0.1102 $\pm$ 0.004 & 0.1184 $\pm$ 0.004  \\ [-1.0ex]
Fe \uppercase\expandafter{\romannumeral10} ~1463.50  & 55.85 & 0.1435 $\pm$ 0.013 & ...     \\ [-1.0ex]
Fe \uppercase\expandafter{\romannumeral11} ~1467.08  & 55.85 & 0.1272 $\pm$ 0.007 & 0.1318 $\pm$ 0.011  \\ [-1.0ex]
Ar \uppercase\expandafter{\romannumeral8 } ~700.26${^\ast}$   & 39.95 & 0.1351 $\pm$ 0.002 & 0.1411 $\pm$ 0.004  \\ [-1.0ex]
Ar \uppercase\expandafter{\romannumeral8 } ~713.82${^\ast}$   & 39.95 & 0.1291 $\pm$ 0.004${^\dagger}$ & ...     \\ [-1.0ex]
Fe \uppercase\expandafter{\romannumeral12} ~1242.00  & 55.85 & 0.1039 $\pm$ 0.004 & 0.1128 $\pm$ 0.004  \\ [-1.0ex]
Fe \uppercase\expandafter{\romannumeral12} ~1349.36  & 55.85 & 0.1078 $\pm$ 0.004 & 0.1177 $\pm$ 0.004  \\ [-1.0ex]
Ca \uppercase\expandafter{\romannumeral10} ~574.00${^\ast}$   & 40.08 & 0.1118 $\pm$ 0.002 & 0.1185 $\pm$ 0.002  \\ [-1.0ex]
Si \uppercase\expandafter{\romannumeral8 } ~1445.75  & 28.08 & 0.1605 $\pm$ 0.004${^\dagger}$ & 0.1674 $\pm$ 0.004${^\dagger}$  \\ [-1.0ex]
S  \uppercase\expandafter{\romannumeral10} ~1196.20  & 32.06 & 0.1250 $\pm$ 0.004${^\dagger}$ & 0.1266 $\pm$ 0.004${^\dagger}$  \\ [-1.0ex]
N  \uppercase\expandafter{\romannumeral5 } ~1238.81  & 14.01 & 0.1709 $\pm$ 0.004 & 0.1804 $\pm$ 0.004  \\ [-1.0ex]
O  \uppercase\expandafter{\romannumeral6 } ~1031.93  & 16.00 & 0.1426 $\pm$ 0.004${^\dagger}$ & 0.1406 $\pm$ 0.004${^\dagger}$  \\ [-1.0ex]
Mg \uppercase\expandafter{\romannumeral9 } ~706.05${^\ast}$   & 24.31 & 0.1669 $\pm$ 0.002 & 0.1795 $\pm$ 0.002  \\ [-1.0ex]
Na \uppercase\expandafter{\romannumeral9 } ~681.72${^\ast}$   & 22.99 & 0.1587 $\pm$ 0.002 & 0.1637 $\pm$ 0.002  \\ [-1.0ex]
Si \uppercase\expandafter{\romannumeral11} ~580.91${^\ast}$   & 28.08 & 0.1271 $\pm$ 0.002 & 0.1204 $\pm$ 0.002  \\ [-1.0ex]
Si \uppercase\expandafter{\romannumeral11} ~604.15${^\ast}$   & 28.08 & 0.1229 $\pm$ 0.005${^\dagger}$ & 0.1313 $\pm$ 0.005${^\dagger}$ \\[-1.0ex]
Mg \uppercase\expandafter{\romannumeral10} ~624.94${^\ast}$   & 24.31 & 0.1485 $\pm$ 0.002 & 0.1524 $\pm$ 0.002  \\ [-1.0ex]
\\ [-2.80ex]
\hline
\end{tabular}
\begin{tablenotes}
\item[a] $^a$ Wavelengths in angstroms were taken from Feldman et al. (\cite{fel97p95}). Lines denoted by an asterisk were observed in second order.
\item[b] $^b$ Half-width at $1/\sqrt{e}$. Instrumental width was removed. The dagger indicates the line widths with no correction for the stray light which is not necessary or possible.
\end{tablenotes}
\label{tab 1}
\end{table*}

In this paper, we investigate the observational data of line widths and find strong correlation between the line widths and ion masses. We present the data used in this paper as shown in Table 1 \citep[][from their Table 3]{dol08p71}. The data comprise two parts. One is for observation at the height of 57$''$ above the solar limb and the other at the height of 102$''$. From Table 1, one can see that the data include 17 (15) emission lines for the observation at the height of 57$''$ (102$''$). Considering that the mass of ion is represented by the mass of the corresponding atom, the masses of the ion species range from $14.01M_H$ to $55.85M_H$, where $M_H$ is the mass of the hydrogen atom. The line width refers to the half-width at $1/\sqrt{e}$ provided through a least-square Gaussian fit including a second-order polynomial. The instrumental width has been removed from line widths. In addition, the effect of possible stray light, i.e. part of radiation that is emitted on-disk and is finally detected off-limb, may be not negligible for off-limb observation \citep{dol03p51,dol04p91}. Correction from the instrumental stray light also has been considered if it is necessary or possible. Finally, it should be noted that some lines, marked with an asterisk in Table 1, were observed in second order, meaning that the measured widths are twice as large as the width that would be measured in first order, which were considered in the study.

\begin{figure*}
\centering
\includegraphics[width=15cm,clip=]{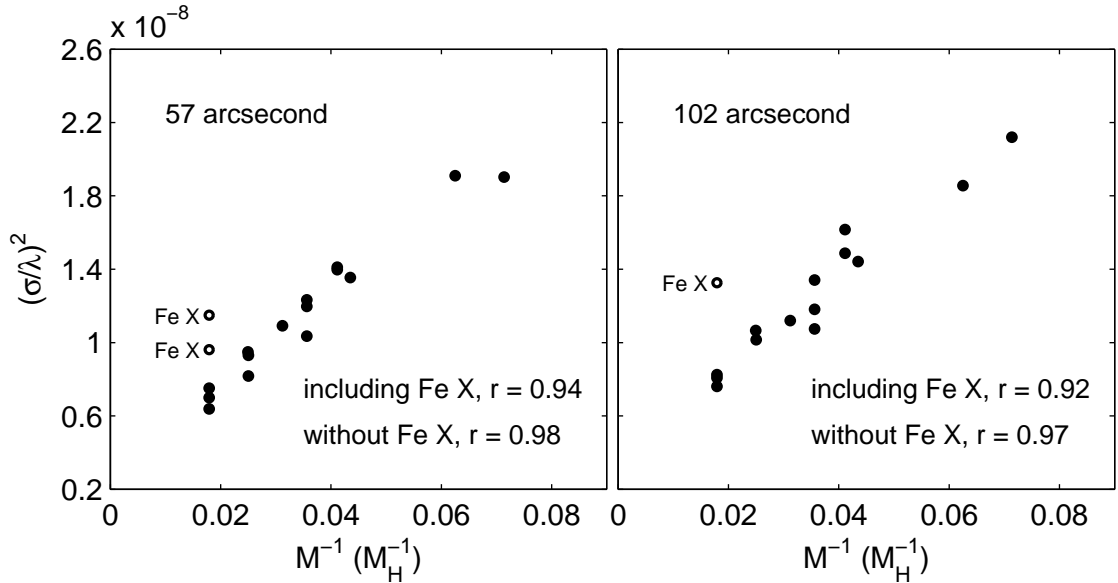}
\caption{Scatter plots of $(\sigma/\lambda)^2$ versus $M^{-1}$ for the data of observations at heights of 57 arcsecond (left panel) and 102 arcsecond (right panel) above the solar limb. The open circles indicate the results for Fe X and the filled circles do for others in both panels.}
\label{fig1}
\end{figure*}

\section{The strong correlation between the line widths and ion masses}

 To proceed with the discussion, we introduce several physical quantities of $\sigma$, $\lambda$ and $M$, referring to the half-width of the observed line profile at $1/\sqrt{e}$, the wavelength and the ion mass, respectively. The left panel of Figure 1 shows a scatter plot of $(\sigma/\lambda)^2$ versus $M^{-1}$ for the data of observation at height of 57$''$, where data points, including filled and open circles, are for the line widths without consideration of the measurement errors, i.e., without error bars. The data points distributed from left to right in the panel correspond to the ions of Fe, Ca, Ar, S, Si, Mg, Na, O, N, respectively. (Note that the ions of Ca and Ar nearly have the same masses.) One can see that these exists a trend of positive correlation between $(\sigma/\lambda)^2$ and $M^{-1}$, especially without the two open circles related to two Fe X lines. For the purpose of evaluating the degree of correlation, we use the Pearson product-moment correlation coefficient which is widely applied as a measure of the degree of linear correlation between two variables \citep{pea03p57}. The value of this correlation coefficient, denoted by $r$, is in the range from $-1$ to 1, where 1 represents total positive correlation, $0$ no correlation, and $-1$ total negative correlation. As shown in the left panel of Figure 1, the correlation coefficient is up to 0.94 including Fe X, and 0.98 without Fe X. Similarly, the right panel of Figure 1 shows the result but for the observations at height of 102$''$. One can find that the correlation coefficient is 0.92 including Fe X, and 0.97 without Fe X, respectively.

\section{The ion temperature and non-thermal velocity}

\begin{figure*}
\centering
\includegraphics[width=15cm,clip=]{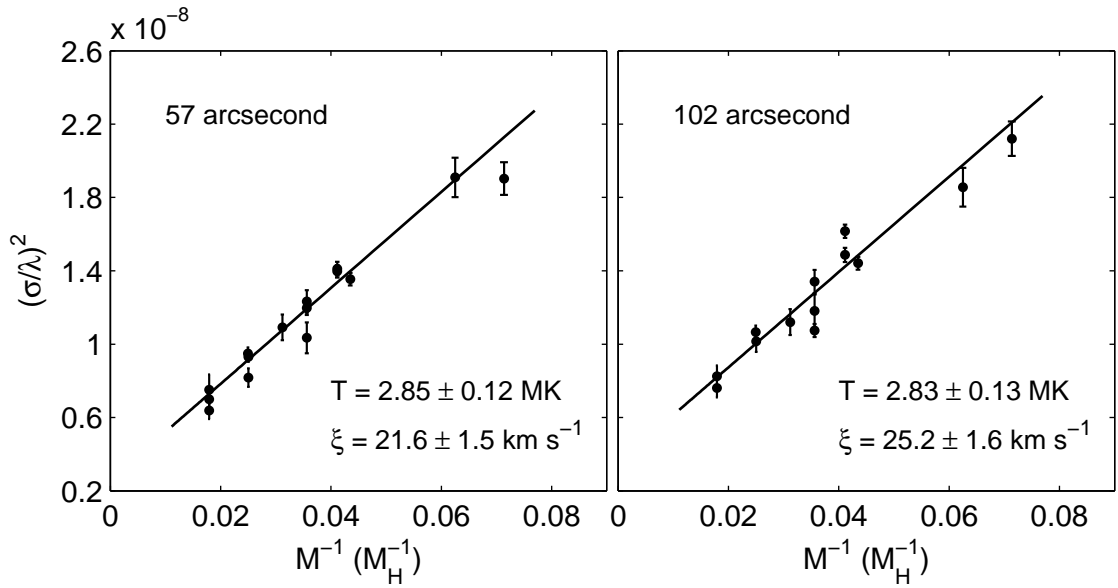}
\caption{The ion temperatures and non-thermal velocities obtained by linear least-square fit, for the data of observations at heights of 57 arcsecond (left panel) and 102 arcsecond (right panel) above the solar limb. The solid lines are the lines of fit to the data. The standard deviations of the both parameters correspond to the measurement errors of line widths shown by error bars in the figure.}
\label{fig2}
\end{figure*}

Now we try to reveal the implication of the strong correlation between the line widths and ion masses. As described in Section 1, both the ion thermal motion and turbulent plasma motion can contribute to Doppler broadening of an emission line. If one assume that each of both the contributing factors has a Gaussian distribution, the half-width of the observed line profile at $1/\sqrt{e}$, i.e. $\sigma$, will follow the formula \citep{der93p17}
\begin{eqnarray}
\sigma^2=\frac{\lambda^2}{2c^2}(\frac{2kT}{M}+\xi^2),
\end{eqnarray}
where $c$ is the speed of light, $k$ the Boltzmann constant, $T$ the ion temperature, $\xi$ the most probable non-thermal velocity. The above equation can also be rewritten conveniently as
\begin{eqnarray}
(\sigma/{\lambda})^2 =aM^{-1}+b,
\end{eqnarray}
with $a=kT/c^2$ and $b=\xi^2/{2c^2}$.

The value of $r>0.9$ derived in Section 3 implies that a linear equation exists to describe the relationship between two variables, i.e. $(\sigma/\lambda)^2$ and $M^{-1}$ in this paper. Based on Equation (2), one may find that the results in Section 3 tend to suggest a point. It is the ions from the same height of polar corona have a common temperature as well as a common non-thermal velocity, so that the values of $a$ and $b$ become constants for different ions. Accepting this suggestion, one can further obtain the ion temperature and non-thermal velocity in terms of the slope $a$ and intercept $b$ of regression line corresponding to Equation (2). The regression line can be obtained by linear least-square fit. For a better fit, we exclude the emission lines for Fe X in the following discussion. Figure 2 presents the results for the data at heights of 57$''$ (left panel) and 102$''$ (right panel), respectively. The measurement errors for each emission line, shown by error bars, are contained in the fit process, leading to uncertainty estimates for the returned parameters. One can find that the deduced ion temperature at height of 57$''$ (102$''$) is $2.85 \pm 0.12$ ($2.83 \pm 0.13$) MK, and the non-thermal velocity is $21.6 \pm 1.5$ ($25.2 \pm 1.6$) km s$^{-1}$. The values of both parameters ($T$ and $\xi$) are compatible with the previous results derived from SUMER observations of coronal hole plasma, i.e., the ion temperatures typically ranging from 1 MK to 4 MK and non-thermal velocities from about 20 km s$^{-1}$ to 50 km s$^{-1}$  \citep{wil12p57}.

\section{Conclusion and discussions}

This paper first reports the finding of the strong correlation between line widths and ion masses. The finding is based on line width observations for the polar corona at heights of 57$''$ and 102$''$. The Pearson product-moment correlation coefficients exceed 0.9, implying a linear relationship between the square of line width relative to the wavelength and the inverse ion mass. The linear relationship, in terms of existing equation (i.e., Equation (2)), suggests that the ions from a given height of polar corona have a common temperature and a common non-thermal velocity. This result can be used to determine the coronal ion temperature and non-thermal velocity. The temperature and non-thermal velocity are obtained by linear least-square fit in this paper. The temperature is around $2.8$ MK at heights of 57$''$ and 102$''$. The non-thermal velocity is 21.6 km s$^{-1}$ at height of 57$''$ and 25.2 km s$^{-1}$ at height of 102$''$, with standard deviations of 1.5 km s$^{-1}$ and 1.6 km s$^{-1}$, respectively.

The finding of strong correlation between line widths and ion masses may give a constraint on the relevant theory. Many authors may question the common ion temperature assumption for corona due to the low plasma density especially for the coronal hole plasma \citep{tuc98p75}, while the results in the present paper are in line with this assumption. With the assumption of a common ion temperature, a lot of studies have been carried out. \citet{see97p87} have derived the coronal ion temperatures and non-thermal velocities in the streamer belt. \citet{dos01p59} investigated the plasmas properties for the case of polar coronal hole. The assumption of a common ion temperature can also be found in the literature \citep{dos00p99,lan03p07,lan06p58,ima09p08}.

On the other hand, one may note that the data points associated with the Fe X lines are not included in doing our linear fit, because these data points make the degree of correlation decreases and seem to be outliers. Some further discussion is desirable. According to the study by \citet{dol08p71}, it was found that the temperature of Fe X is particularly larger than others. A similar result was also obtained by the authors for Fe VIII in the following paper \citep{dol09p51}. The authors suggested that this result constitutes a clue to ion-cyclotron heating, since the study of ion-cyclotron heating shows that the ions with the lowest charge-to-mass ratios always absorb a major portion of the wave energy and leave nothing to appreciably heat the other ions \citep[e.g.,][]{voc02p30}. The present paper, however, do not consider the role of ion-cyclotron heating and assumes that all ions have a common temperature at a given height. Perhaps, it is worthwhile to note that the temperature derived in this paper is in line with that of the ions with high charge-to-mass ratio obtained by \citet[][their Figure 5]{dol09p51}. The derived non-thermal velocities at heights of 57$''$ and 102$''$ are also equal within the error bars in both papers. For instance, the non-thermal velocity at height of 57$''$ obtained by \citet{dol09p51} is $20.5 \pm 2$ km s$^{-1}$ and this velocity is $21.6 \pm 1.5$ km s$^{-1}$ in the present paper. In addition, we should also indicate that, when we keep the data points associated with the Fe X lines in the fit, the value of non-thermal velocity becomes slightly higher, since the intercept of regression line (not shown) is a little larger than that without consideration of the Fe X lines. At the same time, there is a corresponding decrease for the temperature which is determined by the slope of regression line.

Before concluding, two remarks to obtain the present results may be appropriate. First, the data from off-limb observations may be a better choice. Disk observations are subject to a projection effect, which results in commixture of ions with various temperatures and non-thermal velocities. Second, it is also not suitable to directly use the data from off-limb observations summed within a range of heights, since either the ion temperature or non-thermal velocity may be not constant with height due to, for instance, possible energy exchange between ions and waves \citep{cra07p20,wuc13p02}.

Finally, we emphasize that the present study is preliminary and some in-depth study deserves. Most importantly, we have shown that there exists a strong correlation between the line widths and ion masses, at least for the present data. More attention should be paid to this correlation in future studies.

\acknowledgments
Research by D. J. Wu and G. Q. Zhao was supported by the NSFC under grant Nos. 11373070 and 41074107, and by the MSTC under grant No. 2011CB811402. Research by C. B. Wang was supported by the NSFC under grants 41174123 and 41121003, and by the CAS under grant KZCX2-YW-QN512. We are appreciate many fruitful discussions with Professor C. S. Wu during this work. We are also grateful to the anonymous referee for valuable comments.


\end{document}